# Room temperature electroluminescence from isolated colour centres in van der Waals semiconductors

Gyuna Park[1,2†], Ivan Zhigulin[3†], Hoyoung Jung[1,2], Jake Horder[3], Karin Yamamura[3,4], Yerin Han[1,2], Kenji Watanabe[5], Takashi Taniguchi[6], Igor Aharonovich[3,4*], Jonghwan Kim[1,2*]

[1] Center for Van der Waals Quantum Solids, Institute for Basic Science (IBS), Pohang, 37673, Republic of Korea

[2] Department of Materials Science and Engineering, Pohang University of Science and Technology, Pohang, 37673, Republic of Korea

[3] School of Mathematical and Physical Sciences, University of Technology Sydney, Ultimo, New South Wales 2007, Australia

[4] ARC Centre of Excellence for Transformative Meta-Optical Systems, University of Technology Sydney, Ultimo, New South Wales 2007, Australia

[5] Research Center for Electronic and Optical Materials, National Institute for Materials Science, 1-1 Namiki, Tsukuba 305-0044, Japan

[6] Research Center for Materials Nanoarchitectonics, National Institute for Materials Science, 1-1 Namiki, Tsukuba 305-0044, Japan

† These authors contributed equally to this work.

* To whom correspondence should be addressed: I.A. (Igor.Aharonovich@uts.edu.au) and J.K. (jonghwankim@postech.ac.kr)

**Defects in wide bandgap semiconductors have recently emerged as promising candidates for solid-state quantum optical technologies. Electrical excitation of emitters may pave the way to scalable on-chip devices, and therefore is highly sought after. However, most wide band gap materials are not amenable to efficient doping, which in turn poses challenges on efficient electrical excitation and on-chip integration. Here, we demonstrate for the first time room temperature electroluminescence from isolated colour centres in hexagonal boron nitride (hBN). We harness the van der Waals (vdW) structure of two-dimensional materials, and engineer nanoscale devices comprised of graphene - hBN - graphene tunnel junctions. Under an applied bias, charge carriers are injected into hBN, and result in a localised light emission from the hBN colour centres. Remarkably, our devices operate at room temperature and produce robust, narrowband emission spanning a wide spectral range - from the visible to the near infrared. Our work marks an important milestone in van der Waals materials and their promising attributes for integrated quantum technologies and on-chip photonic circuits.**

Colour centres in wide bandgap semiconductors offer a unique opportunity to realise scalable quantum photonic circuitry[1–4]. These "artificial atoms" can often be engineered on demand and exhibit narrowband (~ few nm) emission, even at room temperature. Due to their electronic structure, which involves a ground and an excited state deep within the bandgap of the host matrix, optical excitation is typically employed to probe their fluorescence. However, for many applications, specifically for on-chip quantum devices and scalable quantum circuitry, electrical excitation of emitters is highly sought after[2,4–6].

Electrical excitation of quantum dots[7,8], molecules[9,10] or carbon nanotubes[11] has been thoroughly investigated, but is limited mostly to cryogenic temperatures or challenging to integrate on desired photonic structures. On the other hand, electroluminescence (EL) from solid state defects is only in its infancy. Only a few materials have been explored, including colour centres in diamond[12,13] and silicon carbide[14], as well as selected emitters in zinc oxide[15]. The challenges stem from an efficient engineering of p-type and n-type doping in these materials, along with fabrication of the emitters within the designed circuit[6]. Furthermore, nanoscale engineering of vertical or lateral electrical junctions hosting optically active defects in wide bandgap materials has yet to be demonstrated.

To this end, van der Waals (vdW) materials offer a fascinating opportunity to design a new generation of nanoscale quantum circuitry with electrically driven light sources[16–18]. Hexagonal Boron Nitride (hBN) - the only wide band gap (vdW) crystal - exhibits a unique feature in its ability to host a variety of fluorescent defects[19,20]. Preliminary EL studies from hBN have been attempted, however the emission was mostly confined to the near band-edge at UV spectral range[21]. EL from emitters at the visible and near infrared spectral range, that are highly attractive for quantum applications, have so far also remained elusive. Moreover, all existing attempts were limited to cryogenic temperatures[22], which hindered them from practical room temperature operation.

In this paper, we realise a new generation of vdW devices that exhibit room temperature EL from localised defects in hBN in the visible and near-infrared spectral ranges. Employing vertical tunnel junctions of vdW heterostructures, we achieve efficient electrical excitation of deep defects in hBN. Remarkably, we find the emission is fully polarised and stable over hours of operation. Our work marks a turning point for the future employment of hBN and vdW heterostructures in on-chip, integrated quantum technologies[1–3,5].

Fig. 1a schematically illustrates the device structure where graphene, hBN, and graphene are vertically stacked to form a tunnel junction. Fig. 1b shows an optical microscope image of the representative device. For an emissive layer, we employ carbon-doped hBN crystals (white dashed line labelled with hBN:C) which can host various types of colour centres. hBN:C crystals have various electronic states localised in the atomic scale that can act as colour centres[23,24]. Black dashed lines indicate top and bottom graphene electrodes (labelled as T-Gr and B-Gr, respectively). The heterostructure is encapsulated with high purity hBN, at the top (T-hBN) and bottom (B-hBN). Detailed fabrication procedure is described in the methods section.

Under bias voltage of 24 V, strong and localised EL signals are observed in the wide-field luminescence images (Fig. 1c). The signals are mostly located inside and near the overlapped region of top and bottom graphene electrodes where charge carriers are injected into hBN:C colour centres as schematically described in the inset of Fig. 1c. Due to the ultra-wide bandgap of hBN (~ 6 eV), Fermi level in graphene and band edges of hBN exhibit large energy barriers (over 2 eV) which do not allow carrier injection under low electric fields. Under high electric fields, however, charge carriers can be injected to the band edges of hBN via Fowler-Nordheim tunnelling (Fig. 3c). The mechanism will be described in detail further below.

An example of hBN:C colour centre EL spectrum measured on a confocal microscope is shown in Fig. 1d. We attribute the sharp emission centred around 521 nm to the zero-phonon line (ZPL) transition of an isolated defect in hBN[25]. A Lorentzian fit to the ZPL reveals a narrow full width at half maximum (FWHM) linewidth of 16 meV (3.50 nm). The second prominent sideband feature at 560 nm is the defect's phonon side band (PSB), redshifted from the ZPL by 171 meV, typical of the high energy optical phonon modes in hBN[26]. The details of the spectral analysis can be found in the supplementary section Fig. SI1. Within the same device, other colour centres spanning a range of emission energies were also electrically excited, with selected examples presented in Supplementary Fig. SI2.

To evaluate the stability of fabricated heterostructure devices to electrically excite hBN:C colour centres, and to further examine the emission via its optical properties, we conducted a series of measurements over a span of several days. To begin with, we focus our attention on the colour centre with ZPL positioned at 518 nm, where in Fig. 2 we examine its electrically driven luminescence characteristics and stability. As seen in Fig. 2a, EL spectrum displays two dominant features: a higher energy, narrow peak, which we attribute to a localised acoustically broadened ZPL transition, and a lower energy PSB centred around 564 nm. Fig. 2b shows a colour plot that records the EL spectra over a 1-minute interval over 23-minutes, underscoring the consistency of the luminescence in terms of intensity and wavelength position. The fit parameters of the ZPL line, discerned via a triple Lorentzian fitting approach on the spectra, are elaborated in Fig. 2c-e. The ZPL is described by the first Lorentzian peak of the total fitting function, which displays a centre wavelength within one nanometre fluctuation, with a FWHM value of ~ 5 nm. Additionally, during the measurement, intensity of the ZPL also maintained its brightness, as evident from the consistent EL intensity, which is derived from photon counts.

In order to investigate the electrical excitation mechanism, we measured current-voltage (I-V) characteristics of the devices (Fig 3a). Charge carriers in an electrode can be injected to band edges of a dielectric via Fowler-Nordheim tunnelling process as schematically shown by blue and red arrows in Fig. 3c[27]. For the case of high quality hBN with low impurity density, previous studies[21,28,29] show that I-V characteristics are dominated by Fowler-Nordheim tunnelling of holes at the interface of graphene and hBN (blue arrow in Fig. 3c). Despite the presence of carbon impurities, we find that our device in this study also shows a drastic increase of current from bias voltage of ~ 10 V eventually exceeding 10 μA (orange circles in Fig. 3a). We employ the Fowler-Nordheim tunnelling model for a single injection electrode to analyse I-V characteristics. The tunnelling current satisfies the following equation where $\ln(I/V^2)$ and $1/V$ hold a characteristic linear relationship[27]:

$$\ln(\frac{I}{V^2}) = \ln(\frac{A_{eff}q^3 m}{8\pi h \Phi_B m^* d^2}) - \frac{8\pi\sqrt{2m^*}\Phi_B^{\frac{3}{2}} d}{3hq}\frac{1}{V}$$

$V$ and $d$ represent the bias voltage and insulator thickness, respectively. $A_{eff}$ is the effective tunnelling area, $q$ is the elementary charge, $h$ is the Planck's constant, $m$ is the free electron mass, $m^*$ is the effective mass of the holes in the insulator, and $\Phi_B$ is the tunnelling barrier height for holes at the electrode and the insulator interface. Tunnelling current in our device (orange circles in Fig. 3b) shows an excellent linearity between $\ln(I/V^2)$ and $1/V$ under high bias voltage. The slope is determined by the tunnelling barrier height ($\Phi_B$). According to linear fitting (orange solid line in Fig. 3b), we find nearly identical tunnelling barrier height ($\Phi_B$) to the case of high quality hBN (Fig. SI3). Therefore, we attribute the major mechanism of I-V characteristics in our devices to Fowler-Nordheim tunnelling of holes from graphene to the valence band edge of hBN.

Although injection rate is significantly lower due to larger tunnelling barrier height, Fowler-Nordheim tunnelling can also inject electrons to the conduction band edge of hBN [21,28]. Emergence of EL directly captures small density of electron current via subsequent carrier transport and relaxation processes in hBN as schematically shown in Fig.3c. EL measurement is carried out at a temperature of 10 K to improve sensitivity of the EL signal. Low temperature I-V characteristics (navy circles in Figs. 3a,b) are similar to their room temperature behaviour, consistent with the Fowler-Nordheim tunnelling mechanism which does not depend on temperature. Following tunnelling carrier injection, the radiative recombination process gives rise to EL over a broad spectral range from deep-ultraviolet to near-infrared spectral range. Electrons and holes at the band edges of hBN emit prominent EL at deep-ultraviolet frequencies[21,30] (Fig. 3d). Figs. 3e,f show EL from two representative types of colour centres in our hBN:C crystals. Type 1 in Fig. 3e is attributed to carbon impurities in previous studies[31,32] while type 2 in Fig. 3f is largely unidentified in terms of crystallographic structure. As a function of tunnelling current, EL intensity of type 2 shows strong saturation behaviour (filled squares in the inset of Fig. 3f), which is expected from a single isolated colour centre. On the other hand, saturation behaviour is absent for type 1 (empty square in the inset of Fig. 3f), which can be due to a large density of carbon atoms in our hBN:C crystals ($10^{18} \sim 10^{20}$ atoms/cm$^3$)[23,24]. We note that long-term device operation creates additional defects which modifies I-V characteristics via trap-assisted tunnelling and Poole-Frenkel emission (Fig. SI4).

We now turn to investigate the detailed optical properties of the electrically driven emitters. Figure 4a shows a low resolution EL spectrum across the entire visible spectrum. Due to grating selection, the spectra recorded separately as indicated by the magenta and the navy colours. The measurements are taken at a temperature of 10 K, to provide a better signal to noise ratio of the narrow band emitters observed in EL. It is evident that many colour centres are being electrically excited, which is represented as a multitude of narrow emissive peaks. High resolution EL spectra of selected emitters at the blue, visible and the near-infrared range are shown in Fig. 4b-d. Accounting for acoustic phonon broadening, double Lorentzian fitting functions yielded ZPL positions of 423.51 nm and 625.51 nm with FWHM values of 2.80 nm and 0.95 nm for colour centres in Fig. 4b and Fig. 4c, respectively.

For the particular colour centre at 519 nm, we recorded a full emission polarisation profile, as shown in the inset of Fig 4a. The colour centre exhibits a clear dipole emission profile, even under electrical excitation. The signature of linearly polarised photon emission is a common feature for visible single photon emitters in hBN[33,34]. The visibility of the dipole patterns suggest that each ZPL is indeed from a single colour centre, which is reinforced by the idea of a single electric dipole transition between ground and excited energy states[33,35]. These results also indicate that the optical transition dipole moment is largely in-plane, which is common for most emitter classes in hBN[33,35]. Supplementary material Fig. SI5 provides a more detailed analysis of polarisation measurements, highlighting their significance to the localised nature of the observed colour centres. It also includes additional polarisation plots of two other colour centres with ZPL positions at 506 nm and 620 nm, which also displayed a clear dipole pattern, further reinforcing the localised nature of the electrically driven hBN:C colour centres. Lastly, we examine the optical stability of the 519 nm along with 493 nm and 549 nm colour centres. The measurement is presented as a colour map of spectral acquisitions with over 3 minutes of total duration and can be found in the supplementary information Fig. SI6. We highlight that all three colour centres maintained their ZPL fluctuations below 0.7 nm, with FWHM values of 519 nm and 493 nm colour centres remaining less than 2 nm. Additionally, all three colour centres showed no reduction in luminescence intensity over the span of the whole measurement.

To summarise, we engineered a new vdW heterostructure consisting of hBN and graphene layers that exhibit room temperature, stable EL from isolated colour centres in hBN. Remarkably, the emission spans the visible and the near-infrared spectral ranges, that are areas of interest for quantum technologies employing room temperature single photon sources[4]. We carefully analysed the device performance by investigating electrical current and luminescence as a function of voltages and assigned the major electrical excitation mechanism to Fowler-Nordheim tunnelling injections of electrons and holes. Our results usher in a new era in electrically driven vdW devices for on-chip quantum technologies[3], and propel hBN as a material of choice to realise these devices. The vertical heterostructure stacking is also compatible with large scale engineering[3], as well as integration with on-chip photonic waveguides and cavities, to then couple the circuit components.

Figures

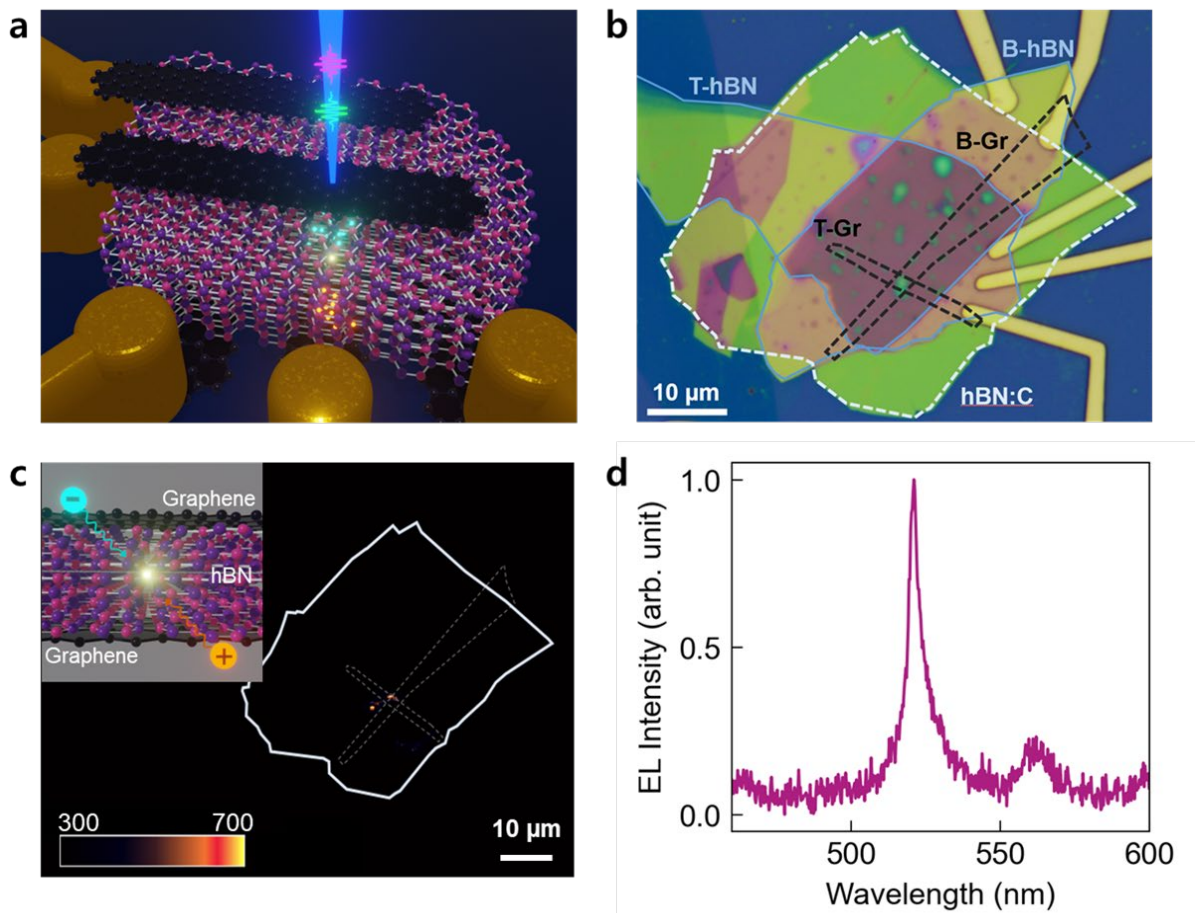

**Figure 1**. **Room temperature EL from isolated colour centres in the van der Waals heterostructure**. **a.** Schematic illustration of the vdW device, An active layer of hBN:C is stacked between top and bottom graphene electrodes. **b**. Optical microscope image of a heterostructure device, hBN:C is highlighted with a white dashed line. Top (T-Gr) and bottom (B-Gr) graphene electrodes are marked with black dashed lines. Blue lines indicate top (T-hBN) and bottom hBN (B-hBN). **c**. Spatial profile of the EL emission, localised inside and near the graphene overlap region. hBN:C is highlighted with white solid line. Grey dashed lines indicate top and bottom graphene electrodes. Inset: Schematic illustration of the electrical excitation mechanism of a colour centre where charge carriers are injected from graphene to hBN via tunnelling process. **d**. Room temperature EL spectrum of a 521 nm colour centre in the vdW device.

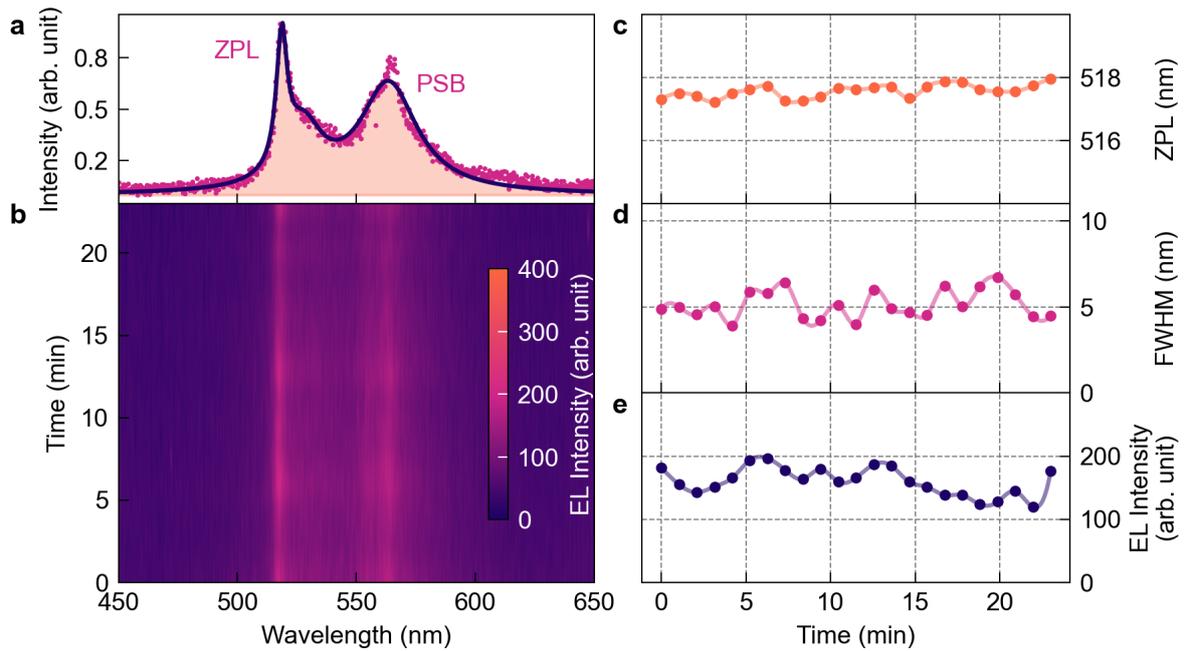

**Figure 2**. **Room temperature EL stability measurement of a colour centre in hBN:C. a.** Normalised luminescence spectra of the colour centre that shows two prominent peaks representative of ZPL and PSB. A triple Lorentzian function yields a ZPL centre wavelength position of 518 nm. **b.** 23-minutes stability measurement presented as a colour map of individual spectra acquired at 60 seconds intervals each. **c-e**. Extracted centre wavelength, FWHM and intensity values, respectively, of the ZPL transition.

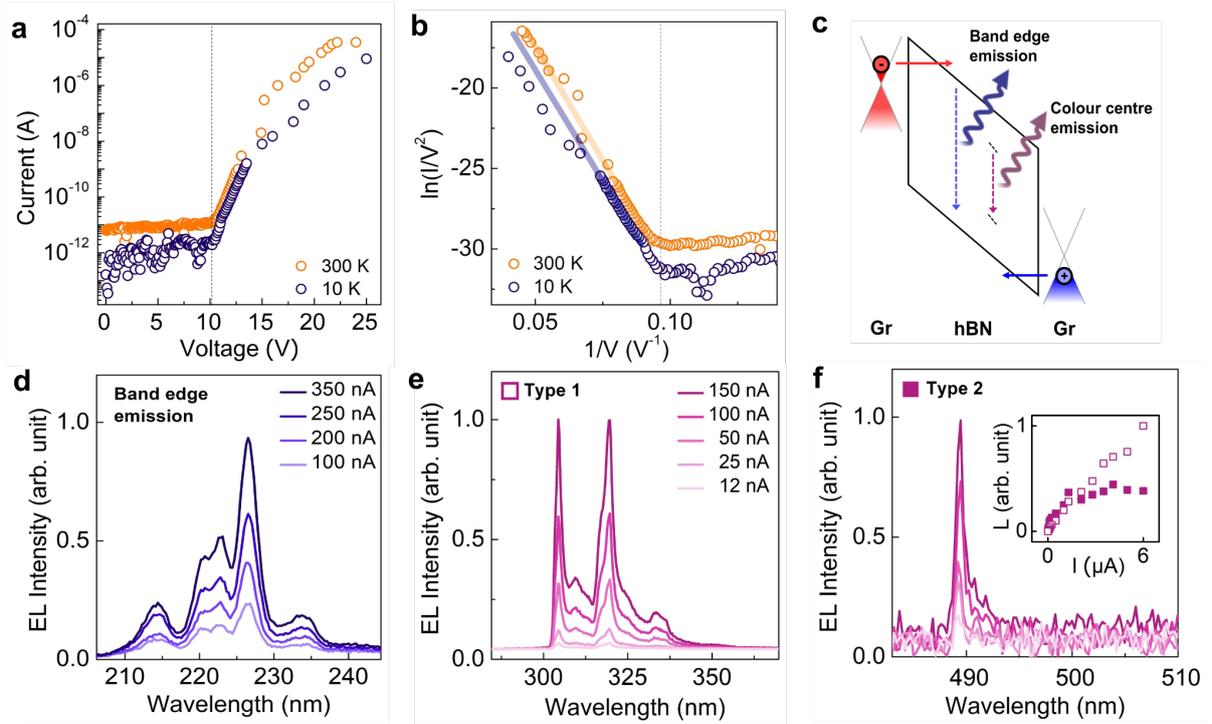

**Figure 3. Electrical excitation mechanism of colour centres in vdW heterostructures. a.** I-V characteristics at temperature of 10 K (navy circles) and 300 K (orange circles). **b.** Fowler-Nordheim plots corresponding to the I-V characteristics in **a**. **c.** A schematic illustration of the electrical excitation mechanism. Band edge and colour centre emissions are marked by navy and magenta dashed lines, respectively. **d.** Deep-ultraviolet EL spectra from band edge EL. **e.** Near-ultraviolet EL spectra from colour centre type 1. **f.** Visible EL spectra from colour centre type 2. The inset shows current-luminescence intensity characteristics of type 1 (magenta empty squares) & type 2 (magenta filled squares) emission. All EL spectra are taken at a temperature of 10 K.

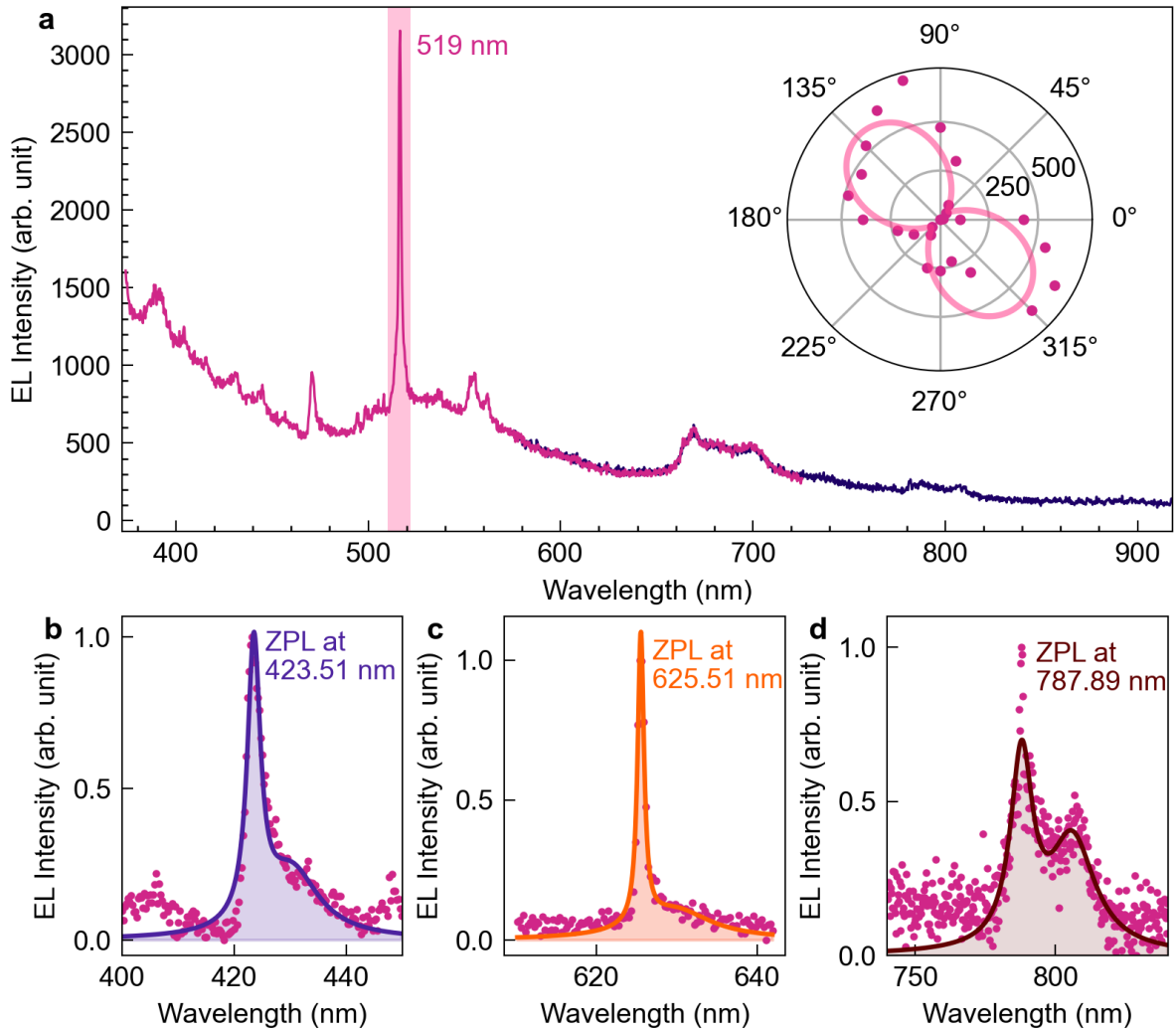

**Figure 4. Optical properties of electrically driven emitters at cryogenic temperature. a.** EL spectra with grating centre wavelength positions at 550 nm (magenta) and 750 nm (navy) reveal a multitude of emission peaks across the visible to near-infrared range. Inset shows emission polarisation measurement of the 519 nm emitter. **b – e.** High resolution EL spectra of isolated colour centres at the blue, red and near infrared spectral range, respectively.